\documentclass[doublecol]{epl2}

\newcommand{\myskip}[1]{}

\renewcommand{\d}{{\rm d}}
\newcommand{\gs}{{\bar g}}
\newcommand{\mg}{{\overline m}_g}

\newcommand{\mx}{m}
\newcommand{\BEQ}{\begin{eqnarray}}
\newcommand{\EEQ}{\end{eqnarray}}
\newcommand{\BEA}{\begin{eqnarray}}
\newcommand{\EEA}{\end{eqnarray}}
\newcommand{\nn}{\nonumber}
\newcommand{\half}{{\frac{1}{2}}}
\newcommand{\Sigmab}{\overline\Sigma}

\newcommand{\br}{{\bf r}}
\newcommand{\bp}{{\bf p}}
\newcommand{\m}{{\rm m}}
\newcommand{\nc}{{\nu{\rm c}}}
\newcommand{\Li}{{\rm Li}}
\newcommand{\s}{{\rm s}}
\newcommand{\p}{\partial}
\newcommand{\fs}{{\rm fs}}
\newcommand{\K}{{\rm K}}

\newcommand{\kpc}{{\rm kpc}}
\newcommand{\Mpc}{{\rm Mpc}}
\newcommand{\eV}{{\rm eV}}
\newcommand{\keV}{{\rm keV}}
\newcommand{\MeV}{{\rm MeV}}

\begin{document}

\title{Do non-relativistic neutrinos constitute the dark matter?}

\shorttitle{Neutrino mass and neutrino dark matter\date{today} }

\author{Theo M. Nieuwenhuizen\footnote{\email{t.m.nieuwenhuizen@uva.nl}}
}

\shortauthor{Th. M. Nieuwenhuizen}

\institute{Institute for Theoretical Physics,  University of
Amsterdam, Valckenierstraat 65, 1018 XE Amsterdam, The
Netherlands}

 {Galaxy groups,
clusters, and superclusters}   {Non-standard-model
neutrinos, right-handed neutrinos, etc.}

\abstract{ The dark matter of the Abell 1689 galaxy cluster is
modeled by thermal, non-relativistic gravitating fermions and its
galaxies and X-ray gas by isothermal distributions. A fit yields a
mass of $h_{70}^{1/2}(12/\gs)^{1/4}$$1.445$ $(30)$ eV. A dark
matter fraction $\Omega_\nu=h_{70}^{-3/2}0.1893$ $(39)$ occurs for
$\gs=12$ degrees of freedom, i. e., for 3 families of left plus
right handed neutrinos with masses $\approx2^{3/4}G_F^{1/2}m_e^2$.
Given a temperature of 0.045 K and a de Broglie length of 0.20 mm,
they establish a quantum structure of several million light years
across, the largest known in the Universe. The virial
$\alpha$-particle temperature of $9.9\pm1.1$ keV$/k_B$ coincides
with the average one of X-rays. The results are compatible with
neutrino genesis, nucleosynthesis and free streaming. The
neutrinos condense on the cluster at redshift $z\sim 28$, thereby
causing reionization of the intracluster gas without assistance of
heavy stars. The baryons are poor tracers of the dark matter
density. }

\maketitle


\section{ Introduction} Dark matter is postulated by Oort to explain
the motion and density of stars perpendicular to the Galactic
plane~\cite{Oort}. Zwicky points out that galaxy clusters must
contain dark matter, ~\cite{Zwicky} while Rubin demonstrates that
galaxies require dark matter in order to explain the rotation
curves of stars and hydrogen clouds~\cite{VRubin}. Nowadays,
gravitational lensing observation is standardized, and dark matter
filaments on the scale of clusters of galaxies between empty voids
can be inferred~\cite{filament}.

The main dark matter candidates are Massive
Astrophysical Compact Halo Objects (MACHOs), Weakly Interacting
Massive Particles (WIMPs), elementary particles, and further, e.
g., axions. Of the total mass of the universe,
$\Omega_B=0.0227/h^2\approx4.23\%$ consists of baryons, of which a
minor part is luminous, a part is located in gas clouds, and a
part makes up the galactic dark matter as MACHOs
~\cite{Gibson96,Schild96,NGS09}. However, the remaining mass of
the universe is non-baryonic, dark energy and dark matter. The
dark matter, with cosmic fraction $\Omega_D=21.4\pm2.7 \%$
according to WMAP5~\cite{WMAP5}, will be the focus of the present
work, where we assume it to consist of fermionic WIMPs.


\section{Thermal fermion model}
We consider non-interacting dark fermions ($x$) with mass $m$ and
${\bar g}$  degrees of freedom, subject to a spherically symmetric
gravitational potential $U(r)$ and in equilibrium at temperature
$T$. The mass density reads

\BEQ \label{rhoxp=} \rho_x=\!\int\!\frac{\d^3
p}{(2\pi\hbar)^3}\,\frac{\gs\mx }{\exp[(p^2/2\mx+mU(r)-\mu)/k_BT]
+1},\EEQ

\noindent where ${\bf p}$ is the momentum and $\mu=\alpha k_BT$
the chemical potential. The spherically symmetric potential
normalized at $U(0)=0$ reads
$U(r)=G\int\d^3r'\rho(r')({1}/{r'}-{1}/{|{\bf r}-{\bf r}'|})$.  It
satisfies the Poisson equation $\label{Ud=} U''+2U'/r=4\pi G\rho.$
In dimensionless variables, $x=r/R_\ast $, $\phi=mU/k_BT$, one has

\BEQ\label{rhox=} \rho_x(r)=- \frac{\gs  \mx }{\lambda_T^3} \,
\Li_{3/2}\left(- e^{\alpha-\phi(x)}\right), \EEQ

\noindent with the polylogarithm Li$_\gamma(z)\!=
\!\sum_{k=1}^\infty \!z^k/k^\gamma$ for $|z|<1$ and analytically
continued elsewhere. The thermal wavelength $\lambda_T$ and a
characteristic scale $R_\ast$ read, respectively,

\BEQ \label{lam=,Rst=} \lambda_T=\left(\frac{2\pi\hbar^2}{\mx
k_BT}\right)^{1/2},\qquad R_\ast=\left(\frac{\lambda_T^3k_BT}{4\pi
\gs G\mx^2}\right)^{1/2}. \EEQ

\noindent Accounting also for the galaxies ($G$) and the hot X-ray
gas ($g$), see below, the Poisson equation will take the form

\BEA \label{Poisson}
 \phi''+\frac{2}{x}\phi'=-\Li_{3/2}\left(- e^{\alpha-\phi}\right)
+ e^{\alpha_G-\bar\beta_G\phi}+e^{\alpha_g-\bar\beta_g\phi}.\EEA

\noindent For $\alpha\to-\infty$ it reduces to a two component
isothermal model, that describes galaxy rotation curves well; for
$\alpha_G,\alpha_g\to-\infty$ it becomes the fermionic isothermal
model.

The total mass inside a sphere of radius $r=R_\ast x$ reads

\BEA \label{Mr} M(r)=\frac{k_BTR_\ast}{ G\mx }\, x^2\phi'(x). \EEA


\section{ Abell 1689}
This is the best studied galaxy cluster, known for its large
lensing arcs, notably one at the Einstein radius $r_E=50''$. It is
well relaxed and spherically symmetric,  with an intruding
subcluster in the North-East, which does not affect the South-West
hemisphere.  It offers a test for the above. From gravitational
lensing the azimuthally averaged mass profile is deduced,
~\cite{TysonFischer}

\BEQ \label{Sigdef}\Sigma(r_\perp)=\int_{-\infty}^\infty\d z
\rho\left(\sqrt{r_\perp^2+z{}^2}\,\right). \EEQ The average
redshift is $z=0.183$. Observations are presented up to
$r_m=h^{-1}$ Mpc, with Hubble constant $H_0=100 h$ km/s Mpc.
Relatively small fluctuations occur in the contrast function
$\Delta\Sigmab(r)$ of the averaged $\Sigma$ within $r$ versus
outside $r$. The first piece reads

 \BEQ\label{Sigav0r} \overline\Sigma(r)=\frac{1}{\pi r^2}
 \int_0^r\d r_\perp 2\pi r_\perp\,\Sigma(r_\perp)\equiv
 \frac{1}{\pi r^2}M_{2D}(r). \EEQ

\noindent The average between $r$ and $r_m$ is directly related to
this, $\Sigmab(r\!\to\!r_m)=[M_{2D}(r_m)-M_{2D}(r)]/
[\pi(r_m^2-r^2)]$, so

\BEQ \label{DeltaSigmab}
\Delta\Sigmab(r) \equiv
\Sigmab(r)-\Sigmab(r\!\to\!r_m)
 = \frac{\Sigmab(r)-\Sigmab(r_m)}{1-r^2/r_m^2}. \EEQ

From (\ref{Sigav0r}) and (\ref{Sigdef}) one derives

\BEQ \overline\Sigma(r)=\frac{4}{r^2}\int_0^\infty\d
r'\,r'\rho(r')
\left[r'-\Re\left(\sqrt{r'^2-r^2}\right)\right].\EEQ After
defining an amplitude $ A\equiv{\hbar^6}/{2\gs^2G^3m^8R_\ast^5},$
eliminating $T$ via (\ref{lam=,Rst=}),
$T={\pi\hbar^6}/{2\gs^2G^2k_B\mx^7R_\ast^4}$, and using
(\ref{Poisson}), also $\Sigmab$  can be expressed in terms of
$\phi'$,

\BEQ \label{Sig=A} \Sigmab(r)=
A\,\Phi\!\left(\frac{r}{R_\ast}\right),\quad \!\!
\Phi(x)=\int_0^\infty\!\d s\,\phi'(x\cosh s).\EEQ

\begin{figure}
\includegraphics[width=8cm]{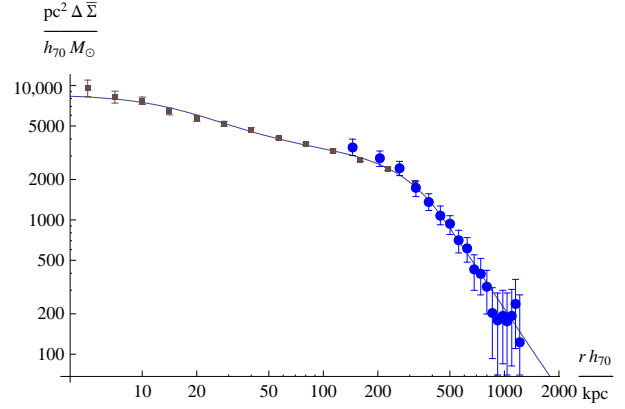}
\caption{ The mass contrast $\Delta \Sigmab$ as function of radius
$r$. Large data points from Ref.~\cite{TysonFischer}, small ones
(at radii $5\, 2^{n/2}h_{70}^{-1}$ kpc with $n=0,..,12$)
constructed from Fig. 6 of Ref. \cite{Limousin}. Full line: The
theoretical profile; it softens below 7 kpc. }
\end{figure}

To proceed, we consider the baryonic matter, galaxies ($G$) and,
mostly, a hot X-ray gas ($g$). In hydrostatic equilibrium,
 $p_i'/\rho_i=-GM(r)/r^2$, ($i=x,G,g$),
both the classical galaxies and the low density gas have a
Boltzmann distribution, $\rho_G\sim\exp(-U/\sigma_v^2)$ with
$\sigma_v$ the line-of-sight velocity dispersion and
$\rho_g\sim\exp(-\beta_g\mg U)$ with $\mg=0.609m_N$ the average
mass in the gas.\footnote{For typical $0.3$ solar metallicity one
has $n_{\rm H}=10\,n_{\rm He}$, so $p_g=2.3\,n_{\rm H}k_BT_g$,
$\rho_g=1.4\,n_{\rm H}m_N$ and $\mg/m_N=1.4/2.3$.} Hydrostatic
equilibrium then imposes $\beta_g=1/k_BT_g$ where $T_g$ is the gas
temperature. With $\bar\beta_G=k_BT/m\sigma_v^2$ and
$\bar\beta_g=\mg T/mT_g$, the densities may be written as
$\rho_i\equiv\gs m\lambda_T^{-3}\exp(\alpha_i-\bar\beta_i\phi)$,
($i=G,g$). This leads to the last two terms of Eq.
(\ref{Poisson}).

We can now make a $\chi^2$ fit of Eqs. (\ref{DeltaSigmab}) and
(\ref{Sig=A}) to the 19 data points of ~\cite{TysonFischer},
combined with 13 points constructed from recent core values for
$M_{2D}(r)$ ~\cite{Limousin} and $\overline\Sigma(r_m)$ from our
model. As seen in Fig. 1, the relative errors increase strongly
with $r$, which reduces the effective number of points. The errors
become more equal if we consider the $\chi^2$ of
$\sqrt{\Delta\Sigmab}$. As explained later, we take
$\bar\beta_g=0.153$ and $\alpha_g=2.36$ at $\bar\beta_G=1$. There
is a minimum $\chi^2=13.617$ at $\bar\beta_G=0.80$. This is close
to the virialized value $\bar\beta_G=1$, so we stick to that with
its $\chi^2=13.645$. With $h\equiv 0.70\,h_{70}$, the correlation
matrix for the upper errors yields

\BEA && A=59.4\pm9.6\,h_{70}M_\odot{\rm pc}^{-2},\quad
\alpha=38.4\pm 3.1,\quad \nn \\
&& R_\ast=297\pm10 \,h_{70}^{-1}\kpc,\quad
\alpha_G=8.26\pm0.32.\EEA

\noindent We present its fit in Figure 1. The WIMP mass reads

\BEQ m=\frac{1}{2^{1/8}\gs^{1/4}}\,
\frac{\hbar^{3/4}}{G^{3/8}A^{1/8}R_\ast^{5/8}}. \EEQ Since
$AR_\ast^5=136 \pm 25 \,h_{70}^{-4}M_\odot{\rm Gpc}^3$, $m$ has a
2\% error,

 \BEA\label{mx=}
\mx&=&h_{70}^{1/2}\left(\frac{12}{\gs}\right)^{1/4}
1.455\pm0.030\,\,\eV. \EEA

\noindent With $T_{\gamma0}=2.725$ K, the global fermion density
is

\BEQ \label{nF=}
n_F=g\frac{3}{4}\frac{4}{11}\frac{\zeta(3)}{\pi^2}
\left(\frac{k_BT_{\gamma0}}{\hbar c}\right)^3 =g\, 55.977 \,{\rm
cc}^{-1}.\qquad \EEQ

\noindent While $\gs$ is the number of states that can be filled
in the cluster formation process, $g$ is the filling factor in the
dark matter genesis. The global mass fraction thus reads

\BEQ\label{Omx=}
 \Omega_x=\frac{n_Fm}{\rho_c}=
\frac{g}{12}\left(\frac{12}{\gs}\right)^{1/4}h_{70}^{-3/2}
0.1893\pm 0.0039. \EEQ

The  gravitino of supersymmetry (s=3/2, $\gs=8$) can explain the
Abell data, but it decouples early, in the presence of
$g_\ast\sim100$ relativistic degrees of freedom, so that $g\sim
0.4$ leads to a small $\Omega_x\sim0.8\%$. The same holds for
other early decouplers~\cite{KolbTurner}. Bosons, like the axion,
can not fit the data because of the tilt in Fig. 1 at $r<200$ kpc;
axionic Bose-Einstein condensation can only exist up to the small
scale $\sqrt{\lambda_TR_\ast}$, which is of no help.

\section{ Neutrinos} They can occupy in the cluster formation
process all $\gs=12$ left and righthanded states, which gives
$m=1.455(30)\,h_{70}^{1/2}$ eV. Neutrinos oscillate,~\cite{nu-osc}
$\Delta m^2_{12}=  8.0^{+0.4}_{-0.3}10^{-5}\eV^2$, $\Delta
m^2_{23}= 1.9 - 3.0\,10^{-3}\eV^2$ so the mass eigenvalues differ.
It is natural to suppose that all virial speeds are equal, so
$T_i=Tm_i / m$. If also their chemical potentials behave as
$\mu_i=k_BT_i\alpha$, the above approach still applies with
$m=(m_1+m_2+m_3)/3$. The degeneracy parameter will change
negligibly,

\BEQ  \gs=\sum_{i=1}^{12} \frac{m_i^4}{ m^4}=12+{\cal O}
\left(\frac{(\Delta m_{23}^2)^2}{m^4} \right).\EEQ The
oscillations bring a shift smaller than our error bars,

\BEQ m_{1,2} =m-\frac{\Delta m_{23}^2}{6m}\mp\frac{\Delta
m_{12}^2}{4m}, \quad m_{3} =m+\frac{\Delta m_{23}^2}{3m}. \EEQ
Sterile masses are expected to weigh keV's or more, see
~\cite{nureview} for a review, but a (near) equality between left
and right handed masses and their abundances is needed to maximize
$\Omega_x$ $\approx $ $h_{70}^{-3/2}\frac{0.19}{12}
\sum_{i=1}^{12}n_im_i/(\frac{1}{12}\sum_{i=1}^{12}n_im_i^4)^{1/4}$.

The de Broglie length in the cluster
$\lambda_{T\nu}={2\gs}{\hbar^{-2}} G\mx^3R_\ast^2 =0.20\,{\rm mm}$
is visible to the human eye; the Compton length is $0.136\,\mu$.
The  A1689 neutrino temperature $ T^A_\nu={\pi G m_\nu A
R_\ast}/{k_B}=0.0447\,\K$ is low and makes them strongly
non-relativistic, with local dispersion

\BEQ \sigma_v^\nu\equiv\left[\frac{\langle{ p_x}^2\rangle}{m^2}
\right]^{1/2}=
\left[\frac{\Li_{5/2}\left(- e^{\alpha-\phi(r/R_\ast)}\right)}
{\Li_{3/2}\left(- e^{\alpha-\phi(r/R_\ast)}\right)}\right]^{1/2}
\sigma_v^G,\EEQ

\noindent which at large $r$ equals the {\it galaxy} velocity
dispersion $\sigma_{v}^G=\sqrt{k_BT/m}=488\pm60$ km/s. The latter
agrees reasonably with estimated speeds in ~\cite{Limousin} and
with the $295\pm40$ km/s of the singular isothermal sphere that
fits the mean galaxy distribution~\cite{Leonard}. Indeed, between
5 and 80 kpc $\rho_G$ looks somewhat like a singular isothermal
distribution.

Neutrinos are abundant, see (\ref{nF=}), but their speed is too
low to leave traces such as Cherenkov radiation. Condensed in
clusters, their local density is large. One has
$n_\nu(0)=-\gs\lambda_T^{-3} \Li_{3/2}(-e^\alpha)=2.3\, 10^8$/cc
and $\rho_\nu(0)c^2=0.34$ GeV/cc, while $n_B(0)=5.1\, 10^9$/cc.
The quantum parameter $N_\nu(r)=-$Li$_{3/2}(-e^{\alpha-\phi})$ is
plotted in Fig. 2. Its maximum is $N_\nu(0)=180$, so quantum
statistics is indispensable. We may define the
quantum-to-classical transition by $N_\nu(r_{qc})=1$. This gives
$r_{qc}= 505\,h_{70}^{-1} \,\kpc$ or diameter $3.3\,10^6$ lyr, a
giant size for quantum behavior.

As seen in Fig. 2, the baryons are poor tracers of the dark matter
density, even they do trace the enclosed mass.

\begin{figure}
\includegraphics[width=8cm]{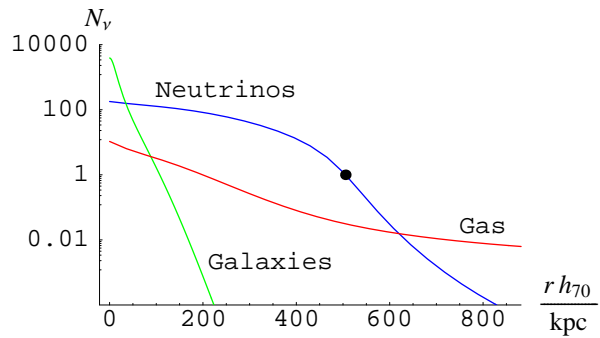}
\caption{Blue line: The neutrino number per cubic thermal length
and per degree of freedom $N_\nu$ as function $r$. The point (505
$h_{70}^{-1}$kpc, 1) separates the region $N_\nu> 1$ which
exhibits strong quantum effects from the classical region
$N_\nu\ll1$. Green line: The galaxy mass density, at the same
scale, is concentrated in the center. Red line: The gas mass
density at the same scale. }
\end{figure}


Neutrino free streaming (fs) in expanding space, $H(t)$ = $\dot
a/a$, is described by a collisionless Boltzmann equation,

\BEQ \label{BEfreeflow}\! \p_t f_\nu+\frac{\bp\cdot\p_\br
f_\nu}{m_\nu} -\left(\bp H+\hat\br\frac{G m_\nu
M(r,t)}{r^2}\right)\cdot\p_\bp f_\nu=0. \EEQ

\noindent Below the Compton temperature $T^C_{\nu}=16,850$ K the
distribution $ f_\nu^\fs={1}/{(e^{p\,c/k_BT_\nu}+1)}$  is long
maintained with sliding $T_\nu=(4/11)^{1/3}T_\gamma$, until the
last, Newtonian term sets in. In the fs regime the scaling $p\sim
k_BT_\nu/c$ implies

\BEQ p_\fs\equiv\sqrt{\langle p^2\rangle_\fs}
=2.567\,\frac{k_BT_\gamma}{c}. \EEQ

\noindent The neutrinos condense ($\nc$) on the cluster when the
typical Newton force $Gm_\nu M(R_\ast)/R_\ast^2=6.56\,
10^{-10}m_\nu\m \s^{-2}$ becomes comparable to the free streaming
force, $p_{\,\fs} H=2.82\,10^{-13}m_\nu (z+1)
\sqrt{\Omega_\Lambda+\Omega_M(z+1)^3}\m \s^{-2}$, which happens
for $\Omega_M=1-\Omega_\Lambda=0.25$ at $z_\nc=28.3$,
$T_\gamma^\nc=77.2\,\K$ and age 120 Myr. Consistency occurs by the
near match of kinetic energies,
$\frac{3}{2}m_\nu[\sigma_v^\nu(R_\ast)]^2=0.64\times (p_{\rm
fs}^\nc)^2/2m_\nu$. The free streaming stops then and the
temperature sinks only moderately, from the cross over value
$T_\nu^A{}^\nc= \frac{3}{2}m_v(\sigma_v^G)^2/k_B=0.13\,\K$ to
present one, $T_\nu^A=0.0447$ K. Literature often mentions 1.95 K
as present neutrino temperature; this derives from relativistic
free streaming, but since they are non-relativistic, the kinetic
temperature would even be lower, $T_\nu^{\rm kin}=2E_{\rm
kin}^{\rm fs}/3k_B=9.6\,10^{-4}\K$. Our $T_\nu^A$ is higher
because the condensation basically stops the cooling. The neutrino
gas then clusters  with a speed of sound $v_s^\nc=\sqrt{5p/3\rho}
= 2620$ km/s at the Jeans scale $L_{J\nu}$ =
$v_s^\nc/\sqrt{G\rho_\nu^\nc}=1.57$ Mpc. The Jeans mass
$M_{J\nu}=\pi\rho_\nu^\nc L_{J\nu}^3/6$ = $1.3\,10^{15}M_\odot$
estimates the total mass (\ref{Mr}) at $r_m=1\,\Mpc/h$, $M_{\rm
tot}(r_m)=6.9\,10^{14}M_\odot$. Only 1.1 \% of this is in
galaxies, and 1.5\% in gas: the A1689 cluster is baryon poor, a
property noticed before ~\cite{TysonFischer,
Andersson}.\footnote{The missing baryons may be located in
intercluster clouds, such as the observed $0.91
\,\keV$ gas
of mass $\sim10^{14}M_\odot$ in a bridge between the
$4.43
\,\keV$ A222 cluster and the $5.31
\,\keV$ A223
cluster~\cite{interclustergas}.}

\section{Virialization, gas profile and X-ray emission}
Partial thermalization takes place since during the condensation
the individual objects, neutrinos, H, He and other gas atoms, and
galaxies move in a time dependent gravitational potential, which
changes each ones energy at a rate proportional to its mass. If
the phase space occupation becomes uniform, a Fermi-Dirac
distribution emerges ~\cite{LyndenBell} and our approach applies;
else it yields an estimate.

This thermalization heats and reionizes the loosely bound parts of
the intracluster gas. The virial proton temperature of the A1689
cluster, $T_p^A$ $=m_pT_\nu^A/m_\nu=2.48$ keV, has the order of
magnitude of the average X-ray temperature $10.5\pm0.1$ keV of the
cluster and the one of the South Western hemisphere,
$11.1\pm0.6\,\keV$ ~\cite{RiemerSorensen}. But virial equilibrium
of the $\alpha$-particles, at temperature $T_\alpha^A=m_\alpha
T_\nu^A/m_\nu=9.9\pm1.1$ keV, explains these observations. For
this reason, we have already taken $\bar\beta_g=\mg/m_\alpha$.

As expected theoretically,  this gives a good match for the mass
profile of the gas, which is deduced from the X-ray profile
~\cite{Andersson}. Fitting their first data point
$M_g=7.79\,h^{-1}M_\odot$ at $30\,h^{-1}\kpc$ to our results sets
$\alpha_g=2.36$ at $h=0.7$, as used above. Then the next 6 data
points, up to $200\,h^{-1}\kpc$, match within the symbol size with
our theory. From there on, it underestimates the data, by a factor
4.5 at the last point at $716\,h^{-1}\kpc$. Reasons for this may
be a decay of the temperature beyond $200h^{-1}\kpc$~\cite{Andersson}
and a contamination of the data from the more relaxed South West
with those from the less relaxed North East.

The bolometric (total) energy emission $1.66\pm0.64\,10^{38}$ W
~\cite{Andersson} must be supplied by a slow contraction of the
cluster. Thus the radiation serves to maintain the virial equilibrium
state as it does for stars~\cite{Balian}.
This will double the A1689 gravitational energy in some 700 Gyr.


\section{ Extended standard model and dark matter fraction}
Righthanded neutrinos have no hypercharge, so they justly do not
enter the $Z$ decay. Their mass is described as for quarks. The
Yukawa eigenvalue $Y_{\nu_1}=2^{3/4}G_F^{1/2} m_{\nu_1}$ can be
expressed in the electron coupling $Y_e=2.935\,10^{-6}$ as
$Y_{\nu_1}$ $= 0.966(20)\,h_{70}^{1/2}Y_e^2$. This exhibits some
order in the lepton masses; in case $h_{70}=1.062(42)$ or
$h=0.744(30)$ it implies

\BEQ m_{\nu_1}=Y_em_e=2^{3/4}G_F^{1/2}m_e^2= 1.4998\,\eV. \EEQ

Active neutrinos (lefthanded neutrinos, righthanded antineutrinos)
have $g=6$ degrees of freedom. Eq. (\ref{Omx=})
leads to a cosmic density $ 0.0952$ $\pm
0.0019\,h_{70}^{-3/2}$, clearly exceeding the $0.028\,h_{70}^{-2}$ of
WMAP5 ($0.013\,h_{70}^{-2}$ when combined with baryon acoustic
oscillations and supernovas) ~\cite{WMAP5}.

The situation can be even more interesting, since the occupation
of the (mostly) righthanded states (sterile neutrinos) can have
become sizeable if there is also a Majorana mass matrix. The
latter couples the neutrino to its charge conjugate, rather than
to the antineutrino, see e. g. ~\cite{GonzalezGarcia}. This allows
neutrinoless double $\beta$-decay, where the two neutrinos emitted
in the $\beta$ decays annihilate each other, and the electrons
leave with opposite momenta. For simplicity we consider the case
of 6 sterile states, bringing the total number of neutrino states
at $\bar g=12$, the case discussed so far. In order to keep nearly
equal masses, we need small Majorana terms $M_i$ ($i=1,2,3$
denotes the families), not the large ones of the see-saw
mechanism. A one-family version of the problem shows that a
(nearly) thermal occupation of sterile modes is possible above the
decoupling temperature $3.5$ MeV, provided $M_i> 3\,
10^{-5}\eV^2/m_\nu\approx 2\, 10^{-5}\eV$ ~\cite{Cline}.
Experimental searches have determined the upper bound $\half
M_e\approx m_{\beta\beta}<0.2-0.7\,\eV$~\cite{2betadecay}. This
filling implies that $g=12$ can be reached, leading to a dark
matter fraction

\BEQ
\label{Omnutot=}
\Omega_\nu\le0.1904\pm 0.0038\,h_{70}^{-3/2}.
\EEQ

The case $g\approx\bar g>12$ is also possible. For $g\sim 33$ it
has enough matter to reach $t_0\approx 1.0/H_0$ without dark
energy.

 \section{Nucleosynthesis}
Our additional relativistic matter can be coded in the enhancement
factor $S$~\cite{Steigman}, $\rho'=S^2\rho$; after $e^+-e^-$
freeze out it reads for $g=12$

\BEQ S= \left(\frac{16+7g
(4/11)^{4/3}}{16+42(4/11)^{4/3}}\right)^{1/2}=1.1854.\EEQ

\noindent This enhances expansion, $H'=SH$, leaving less time for
neutron decay and resulting in too much $^4$He. It can be balanced
by a neutrino asymmetry due to a dimensionless chemical potential
$\xi$, an effect which induces more $n$ decays via $n+\nu_e\to
p+e$,

\BEQ L_e\equiv\frac{n_{\nu_e}-n_{\bar \nu_e}}{n_\gamma}=
\frac{\pi^2}{12\zeta(3)}\left(\frac{T_\nu}{T_\gamma}\right)^3
(\xi+\frac{\xi^3}{\pi^2}). \EEQ

\noindent With $\eta_{10}\equiv 10^{10}{n_B}/{n_\gamma}=121\,
\Omega_Bh_{70}^2$, the cosmic microwave background (CMB) value
reads $\eta_{10}^{\it CMB}=5.60\pm0.15$~\cite{WMAP5}. The $^4$He
value $0.88^{+3.75}_{-0.88}$ does not fit to it, which motivated
to put conservative error bars ~\cite{Steigman}. But the effects
of extra matter and asymmetry on He are large,

\BEQ \eta_{\rm He}=\eta_{10}+100(S-1)-\frac{575}{4}\xi.\EEQ

\noindent So we may fix $\xi$ by matching to $\eta_{10}^{\it
CMB}$. This yields $\xi=0.162$, $ L_e=0.040$. Other authors report
a $^4$He value closer to the one of CMB~\cite{PDGSharkar}, but
this does not modify $\xi$ much. For Li, $\eta_{\rm Li} =
6.05^{+0.13}_{-0.12}$, a similar approach brings~\cite{Steigman}

\BEQ \eta_{10}=\eta_{\rm Li}+3(S-1)+\frac{7}{4}\xi= 6.88^{+0.13}_{
-0.12},\EEQ

\noindent while for deuterium, $\eta_D = 5.92^{+0.30}_{ -0.33}$,
it implies

\BEQ \eta_{10}=\eta_{\rm D}+6(S-1)-\frac{5}{4}\xi
=6.83^{+0.30}_{-0.33}.\EEQ

The freedom in $\xi$ appears to solve the $^4$He discrepancy.
Since the CMB value still has to be rederived for the prior of
neutrino dark matter, the final result may end up near
$\eta_{10}=6.88\pm0.15$, $\Omega_Bh_{70}^2=5.69\pm0.12\%$, to be
compared with $\Omega_Bh_{70}^2=4.63\pm0.12\%$ from WMAP5.
Together with (\ref{Omnutot=}) it would lead to a total matter
fraction $\Omega_M=\Omega_B+\Omega_\nu\approx 24.7\pm0.5\%$, while
WMAP5 reports $25.8\pm3.0\%$.

\section{ Conclusion}
On the basis of three assumptions, Newton's law, quantum
statistics and virialization, we derive the profile of quantum
particles (WIMPs), galaxies and intracluster gas. Because of the
virialization | equal velocity dispersions for WIMPs, galaxies and
$\alpha$-particles, or, more precisely, each ones temperature
proportional to its mass | the WIMPs have a polylogarithmic
profile, while the galaxies and the gas have isothermal profiles.

A fit to total (lensing) mass observations of the cluster Abell
1689 is possible only for fermionic WIMPs with eV mass. The error
of this method is small, 2\% for the present data set. Although we
have not shown the validity of our virial equilibrium assumption,
the explanation of the X-ray temperature of the hot gas, $T_g\sim
10\,\keV=116\,10^6\K$, as the virial temperature of
$\alpha$-particles is striking. Due to collisions that temperature
is shared by the electrons, protons and ions. The predicted mass
profiles for galaxies and gas are also consistent with
observations.

This suggests that our approach cannot be far off, so the WIMP
mass is a few eV and
dark matter is hot. Early decouplers that have been in equilibrium
would yield a small cosmic dark matter fraction~\cite{KolbTurner},
so if they would set the dark matter of the A1689 cluster, there
should also be other dark matter, the major part, but absent in
this cluster, which is unlikely. Therefore early decouplers such
as the gravitino are ruled out as dark matter candidates. The
thermal axion is ruled out because it is a boson.

The case of $\gs=12$ degrees of freedom performs well, pointing at
three families of left and right handed fermions and antifermions.
The obvious candidate is the massive neutrino, because when
condensed in the cluster the left and righthanded states are
equally available. The mass is then $m_\nu=1.445\,h_{70}^{1/2}$ eV
with a 2\% margin and smaller variations between the species due
to neutrino oscillations. There is the striking connection
$m_\nu\sim2^{3/4}G_F^{1/2}m_e^2$. The dark matter fraction of
active neutrinos is then 9.5\%, showing that the cold dark matter
analysis, that allows only 1.3\% at best~\cite{WMAP5}, must
definitively be erroneous.

The scenario in which dark matter is, say, half due to
neutrinos and half due to cold dark matter (CDM) particles
was found viable in
connection with violent relaxation~\cite{Treumann}.
But it does not fit the A1689 cluster.
Indeed, heavy particles have a
Boltzmann distribution. Being
collisionless and relaxed, they are accounted for
already by the isothermal galaxy term in Eq. (\ref{Poisson}) with
$\bar\beta_G=1$. So at best they present 1-3\% of the
A1689 mass, too little for this 50-50 assumption, so it would
again imply the unlikely conclusion that this cluster is
not representative.

The Tremaine-Gunn argument ~\cite{TremaineGunn} of no increase of
the maximal phase space density (except for a factor 2) is
automatically satisfied by the Fermi-Dirac distribution.
\footnote{Let us answer a criticism often met in literature.
Galaxies and dwarf galaxies may have their baryonic dark matter in
the form of MACHOs, H-He planets of earth mass ~\cite{Gibson96},
thousands of which have been observed~\cite{Schild96,NGS09}. In
that scenario the (dwarf) galaxy Tremaine-Gunn bound involves the
{\it proton} mass and is satisfied a million times.}

To describe the dark matter profile of a relaxed galaxy cluster is
a clean problem that involves almost no cosmology, so its daring
predictions pose a firm confrontation to conclusions based on more
intricate cosmological theories, such as the cold dark matter
model with cosmological constant ($\Lambda$CDM model, concordance
model). Indeed, our findings are in sharp contradiction with
present cosmological understanding, where neutrinos are believed
to be ruled out as major dark matter source~\cite{nureview}.
Studies like WMAP5 arrive at bounds of the type
$m_{\nu_e}+m_{\nu_\mu}+m_{\nu_\tau}\le 0.5$ eV. They start from
the CDM paradigm, or from a mixture of CDM and neutrinos, the
reason for this being indirect, namely that without CDM the cosmic
microwave background peaks have found no explanation.
\footnote{CMB peaks may arise without CDM seeds, namely from
viscous instabilities in the baryonic plasma before and at
decoupling ~\cite{Gibson96, NGS09}. Free streaming WIMPs would
then not have time enough to wash out the newly created baryonic
structures.} But the CDM particle has not been detected, so other
paradigms, such as neutrino dark matter, cannot be dismissed at
forehand. The CDM assumption has already
questioned~\cite{Gibson96,Nature} and it is also concluded that
WIMP dark matter has an eV mass ~\cite{Gibson2000}.

The common assumption that light (baryons, Lyman$-\alpha$ forest
aspects) or intracluster gas trace the local dark matter
{\it density} appears to be invalid, even in the absence of a
temperature gradient, for radii at least up to 1.5 Mpc, see fig.
2. Nevertheless, they do trace the enclosed total mass, since this
is imposed by hydrostatic equilibrium. The galaxies ($G$) behave
differently from the neutrinos even for $\bar\beta_G=1$, because
the first ones are classical while the second ones are degenerate
fermions; though non-degenerate, the gas ($g$) behaves differently
from both of them, because it is an ionic mixture with electrons,
implying a parameter $\bar\beta_g<\bar\beta_G$.
This non sequitur nullifies many conclusions in literature,
notably that sterile neutrinos should have keV mass at least and
connections between the Lyman-$\alpha$ forest and local dark
matter densities~\cite{Viel,nureview,Limousin}.

As mentioned, active neutrinos alone would bring about half of the
expected dark matter. In the early Universe the sterile ones can
be created too, at temperatures between $200$ and 3.5\,$\MeV$,
provided neutrinoless double $\beta$-decay is possible, a process
which violates the lepton number. Neutrinos then are Majorana
particles. The related Majorana mass should exceed
$m_{\beta\beta}> 10^{-4}\eV$ or so, which gracefully respects the
experimental upper bound $m_{\beta\beta}<0.2-0.7$
eV~\cite{2betadecay}. Then $\Omega_\nu\le 0.19h_{70}^{-3/2}$.

Both the neutrinos and the antineutrinos fall downwardly in a
gravitational field, as usual.

The neutrinos stream freely, cool in expanding space and condense
on clusters at $z\sim 28$, forming their dark matter. They
simultaneously heat the intracluster gas, which ionizes on its way
to the 10 keV virial equilibrium temperature. This makes it
plausible that it is the neutrino condensation on clusters that,
as a cosmic virial imprint, reionizes all the loosely bound gas,
without any need for heavy stars, the currently assumed cause
~\cite{WeinbergCosmo}.

The central parts of clusters constitute quantum particle
structures of several million light years across. We expect that
the Universe does not contain larger ones, though we make a
reservation for the unknown cause of dark energy.

As usual for clusters, we did not have to invoke a modified Newton
dynamics like MOND~\cite{Milgrom}. The popular NFW mass profile
~\cite{NFW} plays no role, since it deals with heavy,
non-degenerate dark matter, rather than with degenerate fermions.
Fig. 2 shows that the NFW cusp is actually produced by galaxies,
classical objects indeed, but not dark.

The cold dark matter paradigm has to be abandoned, a conclusion
that was already inferred from observed correlations in galaxy
structures ~\cite{Nature}. Hot neutrino dark matter has to be
reconsidered. A gravitational hydrodynamics theory of top -- down
large scale structure formation has been proposed more than a
decade ago~\cite{Gibson96}.

Our theoretical description of virialized galaxy clusters can be
used to process observation data. It can be generalized to
hydrostatic equilibrium with temperature profiles.  Precise
observations of the profiles of the weak lensing, the galaxy
velocity dispersion and the X-rays in the relaxed South West
region of the Abell 1689 galaxy cluster are welcome. The
prediction for the neutrino mass already has a small error and
offers, once it is known, a new way to determine the Hubble
constant. Simulation of the condensation out of free steaming will
test our findings.

As for any physical prediction, the ultimate proof is a direct
observation, in our case of the neutrino mass. The Mainz-Troitsk
tritium $\beta$-decay experiment records the maximal electron
energy in the reaction $^3{\rm H}\to{}^3{\rm He}^+{} +{} e^-{} +{}
\bar\nu_e$, and sets  a bound on the electron-antineutrino mass,
$m_{\bar\nu_e}\le 2$ eV at the 95\% confidence level~\cite{Mainz}.
This leaves our case viable. The KArls\-ruhe TRItium Neutrino
experiment (KATRIN), scheduled for between 2012 and 2015, will
test our prediction by searching for a mass down to 0.2 eV
~\cite{Mainz}, so that our $\sim1.5$ eV regime should be
relatively easy. A next challenge is to settle the Majorana mass
involved in neutrinoless double $\beta$-decay.

The present status of WIMP cold dark matter searches is not good,
since none of the many past or current ones has detected the cold
dark matter particle~\footnote{ An incomplete acronym list of
WIMP/axion dark matter searches and collaborations is: ADMX,
ANAIS, ArDM, ATIC, BPRS, CAST, CDMS, CLEAN, CRESST, CUORE, CYGNUS,
DAMA, DEEP, DRIFT, EDELWEISS, ELEGANTS, EURECA, GENIUS, GERDA,
GEDEON, GLAST, HDMS, IGEX, KIMS, LEP, LHC, LIBRA, LUX, NAIAD,
ORPHEUS, PAMELA, PICASSO, ROSEBUD, SIGN, SIMPLE, UKDM, XENON,
XMASS, ZEPLIN.
}But neither should they still do so if dark matter is just
neutrino matter. The benefit of neutrinos over other dark matter
candidates is that their existence is beyond any doubt. We have
both questioned the reasons to rule them out and derived their
mass. They can be dense, in the Abell 1689 center a billion per 5
cc, but they are non-relativistic and annihilate each other too
rarely to allow observation of the decay products in sky searches.

\acknowledgments We thank Yu. Baryshev, C. Gibson, R. Schild, R.
Balian, J. Smit and J. Kaastra for discussion.


\begin{thebibliography}{0}

\bibitem{Oort} J. H. Oort, Bull. Astron. Inst. Netherl. {\bf VI},
249 (1932).

\bibitem{Zwicky}
F. Zwikcy, Helv. Phys. Acta {\bf 6}, 110 (1933).


\bibitem{VRubin} V. Rubin, N. Thonnard and W. K. Ford, Jr,
Astrophys. J. {\bf 238}, 471 (1980).

\bibitem{filament}
M. J. Geller and J. P. Huchra, Science {\bf 17, 246} 897 (1989).

\bibitem{Gibson96}
C.H. Gibson,  {Appl. Mech. Rev.} {\bf49}, {299} (1996).

\bibitem{Schild96}
R. E. Schild,  {Astrophys. J.} {\bf464}, {125} (1996).

\bibitem{NGS09}
Th. M. Nieuwenhuizen, C.H. Gibson and R. E. Schild, to appear.

\bibitem{WMAP5} E. Komatsu,  et al.,
Astrophys. J. Suppl. {\bf 180}, 330 (2009).

\bibitem{TysonFischer} J. A. Tyson and P. Fischer, Astroph. J.
{\bf 446}, L55 (1995).


\bibitem{Limousin} M. Limousin et al., Astroph. J. {\bf 668},
643 (2007).


\bibitem{KolbTurner} E. W. Kolb and M. S. Turner, {\it The Early
Universe}, (Addison-Wesley, Amsterdam, 1990).

\bibitem{nu-osc}
C. Amsler et al., Phys. Lett. {\bf B667}, 1 (2008).

\bibitem{nureview}
J. Lesgourgues and S. Pastor, Phys. Rep. {\bf 429}, 307 (2006).

\bibitem{Leonard} A. Leonard, et al., Astrophys. J. {\bf 666} 51 (2007).



\bibitem{RiemerSorensen} S. Riemer-S\o rensen et al.,
Astrophys. J. {\bf 693}, 1570 (2009).


\bibitem{Andersson} K. E. Andersson and G. M. Madejski,
Astroph. J. {\bf 607}, 190 (2004). The gas mass profile is part of
Fig. 9.


\bibitem{interclustergas}
N. Werner et al., Astron. and Astrophys. {\bf 482}, L29 (2008).

\bibitem{LyndenBell} D. Lynden-Bell,
Mon. Not. Roy. Astron. Soc. {\bf  136}, 101 (1967).

\bibitem{Balian}
R. Balian and J. P. Blaizot, Am. J. Phys. {\bf 67}, 1189 (1999).

\bibitem{GonzalezGarcia}
M. C. Gonzalez-Garcia and M. Maltoni, Phys. Rep. {\bf 460},1
(2008).



\bibitem{Cline} J. M. Cline, Phys. Rev. Lett. {\bf 68}, 3137 (1992).

\bibitem{2betadecay} H. V. Klapdor-Kleingrothaus et al., Eur. Phys.
J. A{\bf 12}, 147 (2001).

\bibitem{Steigman}
V. Simha and G. Steigman, JCAP {\bf 0808}, 011 (2008).

\bibitem{PDGSharkar}
W. M. Yao et al., J. Phys. G{\bf 33}, 1 (2006).

\bibitem{Treumann}
R. A. Treumann,  A. Kull and H. B\"ohringer,
New J. Phys. {\bf 2}, 11 (2000).


\bibitem{TremaineGunn} S. Tremaine and J. E. Gunn, Phys. Rev. Lett.
{\bf 42}, 407 (1979).





\bibitem{Nature}
M. J. Disney et al., Nature {\bf 455}, 1082 (2008).


\bibitem{Gibson2000}
C. H. Gibson,  {J. Fluids Eng.}{\bf 122}, {830} (2000).

\bibitem{Viel} M. Viel, et al.,
Phys. Rev. Lett. {\bf 97}, 071301 (2006).


\bibitem{WeinbergCosmo} S. Weinberg, {\it Cosmology}, (Oxford,
Oxford, UK, 2008).

\bibitem{Milgrom} M. Milgrom, Astrophys. J. {\bf 270}, 365 (1983).

\bibitem{NFW}
J. F. Navarro, C. S. Frenk and S. D. M. White,
Astrophys. J. {\bf 490}, 493 (1997).


\bibitem{Mainz}
E. W. Otten and C. Weinheimer, Rep. Prog. Phy.s {\bf 71}, 086201
(2008).


\end{thebibliography}
\end{document}